\documentclass[aps,prx,twocolumn,superscriptaddress,showpacs]{revtex4-1}
\usepackage{braket}
\usepackage{xcolor}
\usepackage{mathtools}
\usepackage{hyperref}
\hypersetup{colorlinks=true, urlcolor=blue, citecolor=blue}
\usepackage{times,xspace}
\usepackage{graphicx,graphics,color,epsfig}
\usepackage{amsmath}
\usepackage{amssymb}
\usepackage{amsfonts}
\usepackage{amsfonts}
\usepackage{epstopdf}
\usepackage{bm}
\usepackage{longtable}    
\usepackage{url}           
\usepackage{bm}            
\usepackage{color}
\usepackage{float}
\usepackage[normalem]{ulem}

\begin{document}
	
	\title{Unique $d_{xy}$ Superconducting State in the Cuprate Member Ba$_{2}$CuO$_{3.25}$
	}
	\author{Priyo Adhikary}
    \affiliation{Center for Atomistic Modelling and Materials Design, Indian Institute of Technology Madras, Chennai, 600036, India}
	\affiliation{Department of Physics, Indian Institute Of Technology Madras, Chennai, 600036, India}
	\author{Mayank Gupta}
     \affiliation{Center for Atomistic Modelling and Materials Design, Indian Institute of Technology Madras, Chennai, 600036, India}
	\affiliation{Condensed Matter Theory and Computational Lab, Department of Physics, Indian Institute Of Technology Madras, Chennai, 600036, India}
	\author{Amit Chauhan}
     \affiliation{Center for Atomistic Modelling and Materials Design, Indian Institute of Technology Madras, Chennai, 600036, India}
	\affiliation{Condensed Matter Theory and Computational Lab, Department of Physics, Indian Institute Of Technology Madras, Chennai, 600036, India}
 	\author{Sashi Satpathy}  
	\affiliation{Condensed Matter Theory and Computational Lab, Department of Physics, Indian Institute Of Technology Madras, Chennai, 600036, India}
    \affiliation{Department of Physics \& Astronomy, University of Missouri, Columbia, MO 65211, USA}    
	\author{Shantanu Mukherjee}
	\email{shantanu@iitm.ac.in}
   \affiliation{Center for Atomistic Modelling and Materials Design, Indian Institute of Technology Madras, Chennai, 600036, India}
	\affiliation{Department of Physics, Indian Institute Of Technology Madras, Chennai, 600036, India}
	\author{B. R. K. Nanda}
	\email{nandab@iitm.ac.in}
   \affiliation{Center for Atomistic Modelling and Materials Design, Indian Institute of Technology Madras, Chennai, 600036, India}
	\affiliation{Condensed Matter Theory and Computational Lab, Department of Physics, Indian Institute Of Technology Madras, Chennai, 600036, India}
	\date{\today}
	
	\begin{abstract}
  Recent discovery of superconductivity at a transition temperature of $73$K in the doped layered compound Ba$_{2}$CuO$_{3+x}$ for $x\sim 0.2$ has generated a lot of interest. Experiments in this alternately stacked oxygen octahedral and chain layered structure reveal that a compression of the octahedra causes the Cu- {$d_{z^{2}}$} orbital to lie above the Cu- {$d_{x^{2} -y^{2}}$}  orbital unlike in the well-known cuprate superconducting materials. Our first-principle calculations and low-energy Hamiltonian studies on the $x$ = 0.25 system reveal that this energy ordering results in formation of $d_{z^2}$ dominated electron pockets. 
  The strong nesting in the Fermi pockets leads to an AFM spin fluctuation mediated $d_{xy}$ wave superconducting state dominated by pairing among the $d_{z^{2}}$ orbitals. This is in contrast to the cuprate superconductors (e.g. YBCO) where both electron and hole pockets exist and the superconducting state with B$_{1g}$ symmetry are formed by the $d_{x^2-y^2}$ orbital electrons. Unlike the earlier reports we find the inter-layer hybridization has an important contribution to the low energy band structure and formation of the unconventional superconducting state.  
	\end{abstract}
	\maketitle

A large class of cuprate compounds shows a high-temperature superconducting phase at moderate carrier doping, where Cu $d_{x^{2} -y^{2}}$ orbital electrons are responsible for the formation of the cooper pair condensate.
Recently a new class of overdoped cuprate material \cite{1,2} has emerged, which exhibits higher superconducting transition temperature than the typical cuprates at similar carrier doping \cite{3,41,42}. Among them is the orthorhombic compound Ba$_{2}$CuO$_{3+x}$ \cite{5} with a superconducting transition temperature of $T_c \sim 73$K.  In this material, an octahedral distortion breaks the degeneracy of the $e_{g}$ orbitals, leading to a partially filled $d_{z^{2}}$ orbital and a fully occupied $d_{x^{2} -y^{2}}$ orbital. The presence of high-temperature superconductivity in this material in spite of significantly higher doping levels and low energy physics that is dominated by $d_{z^{2}}$ orbitals provides a new channel for understanding high-temperature superconductivity.
	
		\begin{figure}
		\centering
		\includegraphics[angle=-0.0,origin=0.5,scale=0.34]{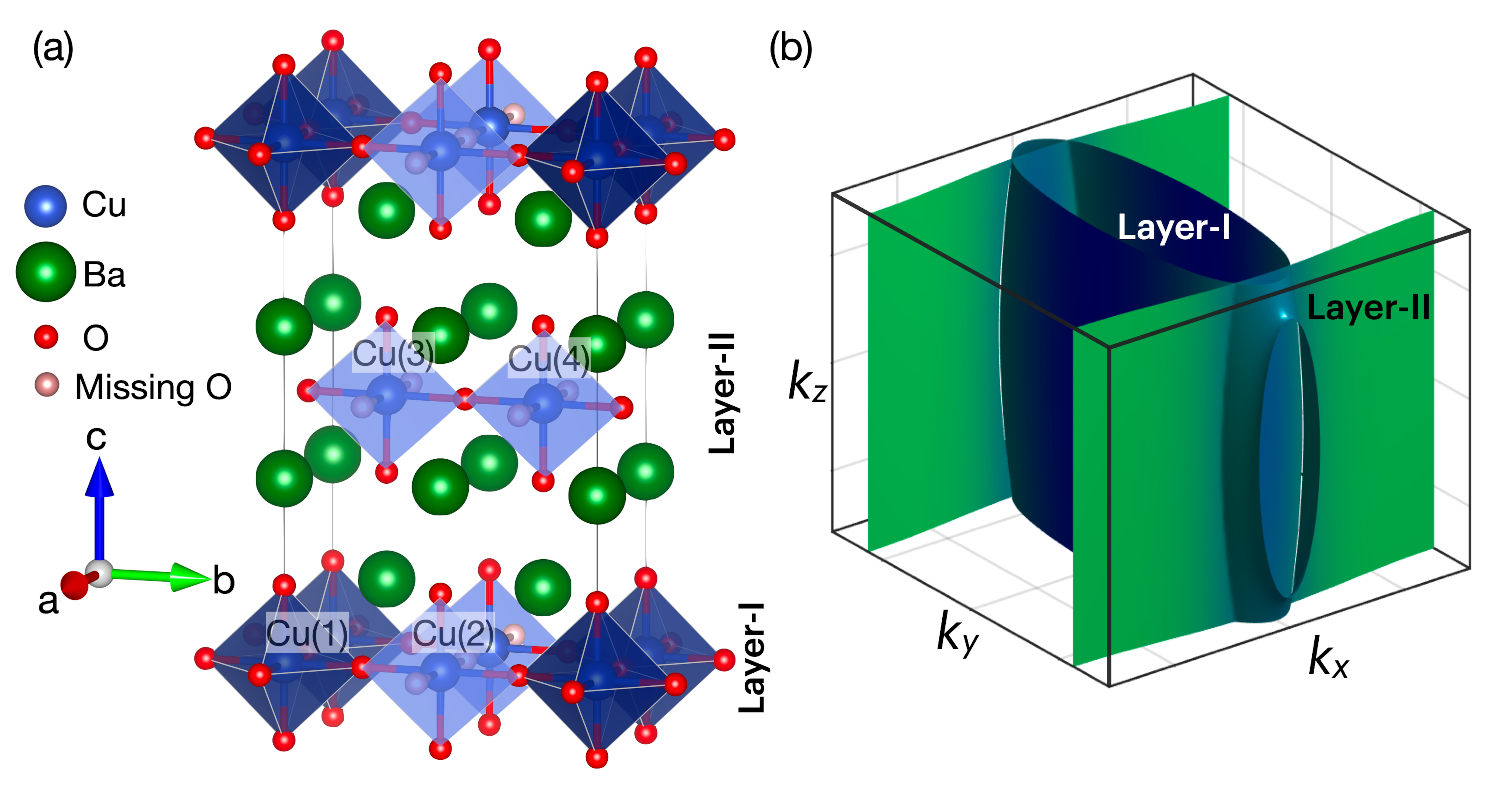}
		\caption{(a)  Crystal structure (Pmmm) of Ba$_{2}$CuO$_{3.25}$ (BCO). Layer-I has alternate stacking of octahedral and square planar Cu-O complexes (along $b$). The octahedra form a corner-share network along $a$-direction. In layer-II, due to missing oxygen atoms when compared to Ba$_{2}$CuO$_{4}$, only square planar complexes exist and they form a chain along b-direction (b) The Fermi surface of BCO. Center elliptical pocket with strong nesting is formed by Cu(1)-$d_{z^2}$ orbital of layer-I while the open electron sheets are formed by  Cu(3, 4)-$d_{b^2-c^2}$ layer-II. 
}
		\label{ref:fig1}
	\end{figure}
Recent experimental studies on Ba$_{2}$CuO$_{3+x}$ with $x$  = 0.2 \cite{5,7} measuring oxygen K-edge X-ray absorption spectra (XAS) estimate 40\%  doping, that is significantly higher than doping in overdoped cuprates like YBCO.
 The Zhang-Rice singlet state is observed at this oxygen doping with  dominant pre-peak at 528 eV photon energy \cite{5}.
	 The XAS measurements \cite{5} on the Ba$_{2}$CuO$_{3.2}$ suggested a compressed octahedral structure. This prompted Maier $et. \hspace{0.1cm}al.$ \cite{8} to suggest a simplified two-band model using  $d_{z^{2}}$ and $d_{x^{2} -y^{2}}$  orbitals within the {\bf 214} structure. They propose two dome superconductivity one at low doping and the other at high oxygen doping. Similar calculations based on spin-fluctuation theory on Lieb lattice \cite{9} structure give $s\pm$ wave superconductivity. 
  
  The unit cell of Ba$_{2}$CuO$_{3.25}$ (BCO) has two layers, layer-I and layer-II (see Fig. \ref{ref:fig1}). The DFT+DMFT based calculations \cite{6} proposed the presence of a charge transfer between layer-I and layer-II and proposed that due to the presence of a quasi-1D band, antiferromagnetic spin-fluctuation may occur to give rise to  superconductivity. The specific-heat measurement \cite{5} on BCO indicates that superconductivity is very anisotropic in contrast to the exponential jump at $T_{c}$ of conventional electron-phonon superconductivity. 

 In this letter, using a combination of first-principles DFT calculations and a spin fluctuation mediated superconducting pairing mechanism \cite{10}, we find a crucial inter-layer hybridization present in BCO which not only helps to stabilize a novel $d_{xy}$ symmetry superconducting state but also leads to a weak 3-dimensional character to the superconducting gap function. 
Experimentally, in the XAS spectra, a transition  $2p^{6}3d^{9}L \rightarrow 2p^{5}3d^{10}L$ is seen \cite{5} which we attribute to this hybridization and is associated with hopping between Cu-$d$ and O-$p_{y}/p_{z}$ orbitals. 
 We identify the ground state superconducting gap function for decoupled individual layers and the bulk BCO that includes inter-layer hybridization. The analysis provides a comparative study between a layer decoupled and hybridized low energy Hamiltonian to elucidate the role of the latter towards superconductivity. Despite the larger electronic doping, we find that the Fermi surface (FS) remains significantly nested, and it leads to a large paramagnetic susceptibility and superconducting pairing potential. The obtained superconducting gap function belongs to pairing between electrons predominantly in $d_{z^{2}}$ orbitals.

 We find that the gap symmetry and the strength of the pairing of the bulk are different from a model incorporating decoupled layers. Furthermore, due to the presence of inter-layer hybridization, the magnitude of pairing eigenfunction has a weak $k_{z}$  dependence, which is otherwise absent in the case of decoupled layers. The pairing symmetry of the bulk has a gap function with $d_{xy}$ symmetry and the pairing symmetry of individual layer-I is $s+$ type and layer-II is $s-$ type with additional nodes at the FS. Further, the broken $C_4$ rotational symmetry due to structural distortion results in a pairing symmetry belonging to the $D_{2h}$ point group. 

	{\it Electronic structure -} Employing the DFT+DMFT method, Worm \textit{et al.} \cite{6} have examined the electronic structure of BCO to make a broad prediction of the presence of an almost half-filled, strongly nested, quasi-1D $d_{b^2-c^2}$ band which is probable cause of superconductivity. Here, we would like to make a comprehensive analysis of the electronic structure using DFT and to develop a tight-binding (TB) model that examines the inter-layer coupling and its effect on FS of the BCO. As discussed later and in the Supplementary Material (SM), the minimal basis set TB model is developed by both the Slater-Koster formulation and the Lo\'wdin downfolding technique. \par
\begin{figure}
		\begin{center}
			\rotatebox{0}{\includegraphics[width=0.48\textwidth]{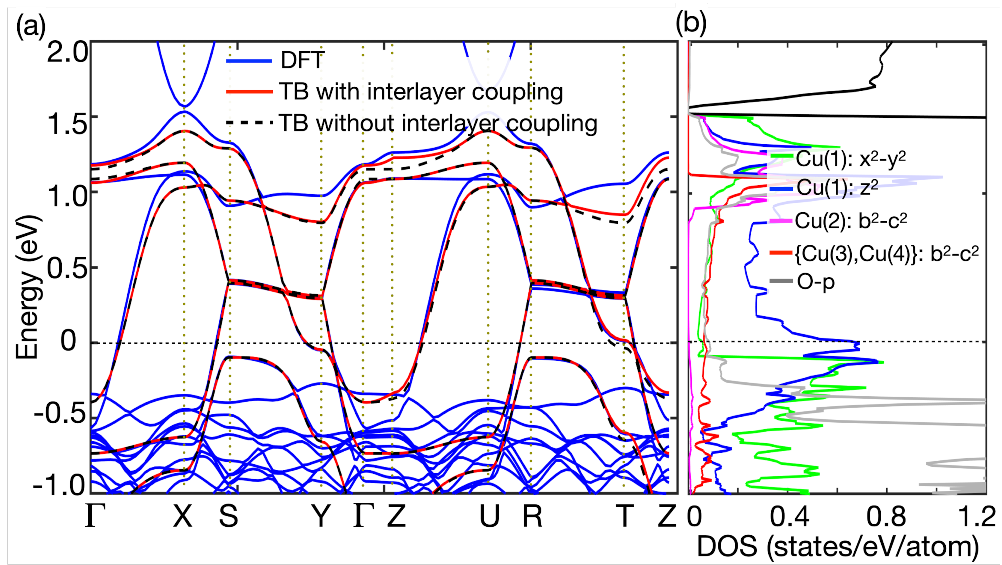}}
		\end{center}
		\caption{ (a) The five-band TB model fitted with DFT bands with and without inter-layer coupling. 
  The inter-layer hybridization shifts the Van Hove singularities near Y and T points at the high symmetry $k$-path of the Brillouin zone. The energy gap between Van Hove singularities is absent for the decoupled individual layers. The high symmetry $k$-path used to plot the band structure is provided in the Fig. S1 of the SM. (b) DFT obtained partial density of states.  }
		\label{ref:fig2}
	\end{figure}

 The DFT-derived bands are shown in blue in Fig. \ref{ref:fig2} (a), along with the  orbital resolved density of states shown in Fig. \ref{ref:fig2} (b).  Details of the DFT calculations and orbital resolved band structure are provided in the SM. Below we mention the main findings of DFT results. 
	(I)	$d_{z^{2}}$  orbital of Cu(1) is within the range of -0.56 eV to 1.13 eV w.r.t. the Fermi level (E$_F$). The $d_{x^{2} -y^{2}}$ orbital is completely occupied and lies in the range -1.57 eV to -0.09 eV. This highlights the role of $d_{z^{2}}$ orbitals in the typical energy scales associated with the superconducting transition.
	(II) $d_{b^{2} -c^{2}}$ orbital of Cu(2) is about 0.89 eV above E$_F$  (0.89 eV to 1.56 eV).
	(III) The O-$p$ states are extended in valence bands and lie up to -0.27 eV below E$_F$. \par

 
	To gain further insight into the electronic structure, specifically to obtain the eigenvectors of the states occupying the Fermi level, we developed a low energy TB Hamiltonian initially with a 14 orbital basis and later downfolded to a five orbital basis. The Hamiltonian reproduces well the DFT band structure in the corresponding energy range (see Fig. \ref{ref:fig2} (a)) and thereby enables us to calculate the random phase approximation (RPA) spin susceptibility.
 
 In this model, four Cu atoms in the unit cell contribute five Cu-$d$ orbitals, and the nearest neighbor O atoms contribute to the nine O-$p$ orbitals near the Fermi level (See SM for detailed analysis). We write the Hamiltonian in the following form,
	\begin{equation}
		H = \sum_{\alpha\beta } \sum_{\substack{{\bf k} \\ \sigma\in(\uparrow,\downarrow)}}\Big[ \xi_{\alpha\beta}({\bf k}) + \mu_{\alpha} \delta_{\alpha\beta}\Big] c_{{\bf k},\alpha,\sigma}^{\dagger}c_{{\bf  k},\beta,\sigma}.
	\end{equation}

 Here,  $c^{\dagger}_{{\bf k},\alpha,\sigma}$ ($c_{{\bf k},\alpha, \sigma}$ )  is the Fermion creation (annihilation) operator for orbital $\alpha$  with spin $\sigma=\uparrow,\downarrow$. $\xi_{\alpha\beta}({\bf k})$ is the kinetic energy term containing hopping parameters and  $\mu_{\alpha}$ is the onsite energy. 
	
	Since there are two weakly coupled inequivalent layers in the system, the TB Hamiltonian contains intra-layer and inter-layer Hamiltonian contributions. In a matrix form, the Hamiltonian can be expressed as:
	\begin{equation}
		H = \left( \begin{array}{cc}
			H_{l1}& H_{l1-l2} \\
			H_{l1-l2}^\dagger & H_{l2}\\
		\end{array} \right)
	\end{equation}
	
	Here, $H_{l1}$ and $H_{l2}$ are the intra-layer Hamiltonian submatrices for layer-I and layer-II respectively, and $H_{l1-l2}$ accounts for the inter-layer hybridization.  Using   Lo\"wdin downfolding \cite{11} procedure, we obtain an effective Hamiltonian by integrating out the oxygen subspace while keeping only the Cu-$d$ orbitals in the Hamiltonian. The matrix elements and the downfolding formalism are provided in the SM. 

 	 	\begin{figure}
		\begin{center}
		\rotatebox{0}{\includegraphics[width=0.45\textwidth]{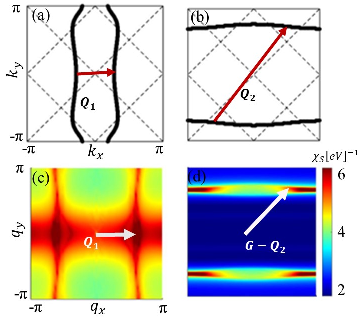}}
		\end{center}
		\caption{(a), (b)  Fermi surfaces of decoupled individual layer-I and layer-II in ${k_x}-{k_y}$ plane. The Fermi pockets are electron-like, coming from the $d_{z^{2}}$ orbital of Cu(1) atom and  $d_{b^{2} -c^{2}}$ orbital of Cu(3) atom. (c), (d)  RPA spin susceptibility [$\rm Tr[\chi_{s}]$] of decoupled individual layer-I and layer-II within the ${q_x}-{q_y}$ plane.  The  dominant nesting vectors are denoted by ${\bf Q_{1}}$, ${\bf Q_{2}}$. ${\bf G}$ is the reciprocal lattice vector. }
		\label{ref:fig6}
	\end{figure}
 
	The TB bands with and without inter-layer hybridization are shown in Fig. ~\ref{ref:fig2} (a) and are compared with the DFT obtained band structure. The lack of inter-layer coupling does not reproduce the band structure well with the subtle differences can be seen in Fig.~\ref{ref:fig2} (a) when momenta changes along $T (0, \pi, \pi)$  to $Z(0, 0, \pi)$ direction.
  The distinction between them comes from  inter-layer hybridizations Cu(1)-$d$ -- Cu(3, 4)-$d$ of strength $t_{12}^{(1)}$ and between Cu(1)-$d$ -- O-$p$ of strength $t_{12}^{(2)}$. We find $t_{12}^{(1)}$ to be one order of magnitude higher than $t_{12}^{(2)}$ (see SM).
  Most importantly, the inter-layer hopping pushes the Van Hove singularity slightly above the Fermi level.  The $d_{z^{2}}$ orbital of Cu(1) and $d_{b^{2} -c^{2}}$ orbital of Cu(3, 4) cross Fermi level and form electron like pockets. Both electron pockets hybridize near  $Y(0, \pi, 0)$ point. The $d_{x^{2} -y^{2}}$ orbital of Cu(1) octahedra lies slightly below the Fermi level. This happens because of octahedral distortion in the BCO structure at very high doping. We can see in Fig. \ref{ref:fig2} (b) that Cu(1)-$d_{z^{2}}$  and Cu(3, 4)-$d_{b^{2} -c^{2}}$ have larger DOS at the E$_F$ than Cu(2)-$d$ orbitals. The FS topology of the bulk is shown in Fig.~\ref{ref:fig1} (b). 
		
	{\it Superconducting state -}
	The multi-orbital superconducting pairing Kernel is derived from a spin fluctuation pairing mechanism.  \cite{41,42,121,122,123,124,125,126,127,128,131,132,133,134,135,141,142,143,144,145,146,151,152} The fluctuation exchange approximation (FLEX) that has been successfully utilized to extract the ground state superconducting states of both cuprate \cite{41,42,121,122,123,124,125,126,127,128} and iron based superconductors \cite{41,42,131,132,133,134,135}. The pairing Kernel involves contribution from paramagnetic and charge susceptibilities that are calculated from a Hubbard-Hund Hamiltonian within the RPA (See SM for detailed analysis). Finally, the pairing interaction is included in the self-consistent linearized gap equation in order to extract the ground-state superconducting gap functions. The gap equation reads, 
 \begin{eqnarray}
	\Delta_{\nu}({\bf k}) &=& -\lambda\frac{1}{\Omega_{\rm BZ}}\sum_{\nu',{\bf q}}\Gamma'_{\nu\nu'}({\bf k,q})\Delta_{\nu'}({\bf k+q}).
	\label{SC2}
\end{eqnarray}
Where $\Gamma'_{\nu\nu'}({\bf k,q})$ is SC pairing potential and $\lambda$ is the pairing strength. We obtain SC pairing potential  by expanding the interaction term of the Hubbard Hamiltonian in a perturbation series and collecting the bubble and ladder diagrams,
	\begin{eqnarray}
		\tilde{\Gamma}_{s}({\bf q})&=&\frac{1}{2}\big[3{\tilde U}_{s}{\tilde \chi}_{s}({\bf q}){\tilde U}_{s} - {\tilde U}_{c}{\tilde \chi}_{c}({\bf q}){\tilde U}_{c} + {\tilde U}_{s}+{\tilde U}_{c}\big].
	\end{eqnarray}
Here, $\chi_{s}(\chi_{c})$  is the RPA spin (charge) susceptibility.

	\begin{figure}
		\begin{center}
			\rotatebox{0}{\includegraphics[width=0.5\textwidth]{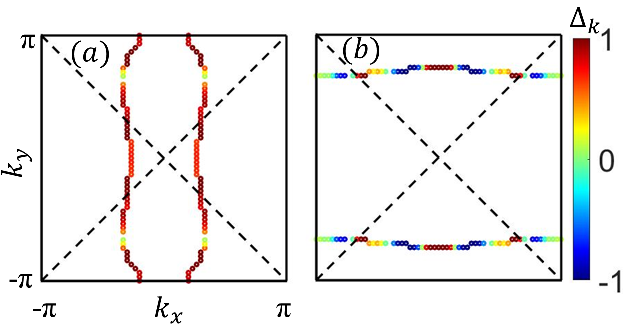}}
		\end{center}
		\caption{(a), (b) The superconducting pairing eigenfunction for the largest eigenvalue on the Fermi surface for decoupled individual layer-I and layer-II respectively. The pairing symmetry of layer-I and
       layer-II belongs to A$_{1g}$ irreducible representation of the D$_{2h}$ point group.}
		\label{ref:scnohyb}
	\end{figure}
	
In BCO, the effect of a broken $C_4$ rotational symmetry can be seen on the FS and corresponding spin susceptibility calculations. We first explore the Hamiltonian in the limit of no inter-layer hybridization. In Figs.~\ref{ref:fig6} (a) and (b) we show the FS topology of layer-I and layer-II respectively at $k_z=0$. We find that layer-I shows a stronger 2D dispersion as compared to layer-II. The effect of the quasi-1D nature of layer-II shows up in the 1D susceptibility peaks observed from its FS  nesting. The dominant FS nesting vectors for individual layers are also shown in Figs.~\ref{ref:fig6} (a) and (b). We show the spin susceptibility of layer-I and layer-II in Figs.~\ref{ref:fig6} (c) and (d) respectively. This chain-like 1D FS enhances spin susceptibility in each layer for small values for Hubbard interactions and leads to a dominant nesting for layer-II at the incommensurate wavevector ${\bf Q_{2}} =(\pm 0.96 \pi, \pm 0.68 \pi)$. Similarly, the spin susceptibility result of layer-I leads to a corresponding peak at ${\bf Q_{1}} =(\pm 0.66\pi, 0)$. We find that $\chi_{s}$ at $\bf Q_{2}$ is larger than the corresponding maximum for $\bf Q_{1}$.

When inter-layer hybridization is absent, SC of each layer is decoupled from the bulk BCO. Peaks of spin susceptibility will determine the maximum pairing potential when nesting condition ($\xi_{{\bf k+Q}} = -\xi_{{\bf k}}$ ) is satisfied by the momentum ${\bf  q= Q}$ at the FS. The strong deviation of the dominant susceptibility from the $C_4$ susceptibility of the well-known cuprates can lead to corresponding deviations in the superconducting state. We plot the superconducting gap function for layer-I and layer-II for the largest pairing eigenvalue in Figs.~\ref{ref:scnohyb} (a) and (b) respectively. Colormap blue to red denotes the sign of the pairing symmetry. The pairing symmetry of layer-I and layer-II leads to a dominant spin singlet superconducting gap that would transform as an A$_{1g}$ irreducible representation of the D$_{2h}$ point group symmetry. Whereas for layer-I the gap is only anisotropic near the region of large curvature of the FS, the gap on layer-II shows the presence of accidental nodes. We also find that the superconducting gaps have a 2D structure over the FS within negligible $k_z$ dispersion.  

 The formation of a SC gap with A$_{1g}$ symmetry despite of a repulsive pairing interaction is explained by the dominance of the inter-orbital pairing channel over the intra-orbital pairing contribution (see the discussion in Section IV of SM). This large off-diagonal contribution leads to an attractive pairing. Such a scenario can also be induced by Hund's interaction \cite{roig}, although for BCO it is already present at the non-interacting level.  


		  	\begin{figure}
		\begin{center}
			\rotatebox{0}{\includegraphics[width=0.5\textwidth]{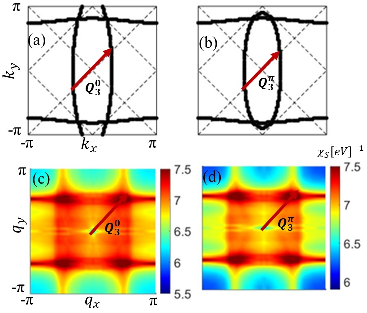}}
		\end{center}
		\caption{(a), (b) The Fermi surface of BCO in the presence of inter-layer hybridization at two different ${k_z}$ values. The weak
3D dispersion induced by the inter-layer hybridization causes  ellipticity of the electron pocket at $k_{z}=\pi$ as compared to the pocket at $k_{z}=0$.  (c), (d) RPA spin susceptibility, $\rm Tr[\chi_{s}]$ for two ${q_z}$ values.  The  dominant nesting vectors are denoted by  ${\bf Q_{3}^0}$, ${\bf Q_{3}^\pi}$. }.
		\label{ref:fig3}
	\end{figure}

We next include inter-layer hybridization in the non-interacting Hamiltonian. In Figs.~\ref{ref:fig3} (a) and (b) we show the FS of bulk BCO at $k_{z}=0$ and $k_{z}= \pi$ respectively. Interestingly, the inter-layer hybridization not only enhances the dispersion of the electronic bands along the $k_z$ direction, it also leads to a significant shift of the dominant susceptibility peak. The RPA spin susceptibility at $q_{z}=0$, and $q_{z}= \pi$ are shown in Figs.~\ref{ref:fig3} (c) and (d). As shown in Fig.~\ref{ref:fig3} (c), the dominant nesting vector is still in the $k_x$-$k_y$ plane but leads to a susceptibility peak at the wave vector ${\bf Q_{3}^0}= (\pm 0.48 \pi,\pm 0.52 \pi)$ for $k_z=0$. The larger ellipticity of the electron pocket at $k_{z}=\pi$ as compared to the pocket at $k_{z}=0$ shows the weak 3D dispersion induced by the inter-layer hybridization, and as shown in Fig.~\ref{ref:fig3} (d) leads to a susceptibility peak at around ${\bf Q_{3}^{\pi}}= (\pm 0.44\pi,\pm 0.56\pi)$ wavevector. 

\begin{figure}[H]
 	\begin{center}
 		\rotatebox{0}{\includegraphics[width=0.5\textwidth]{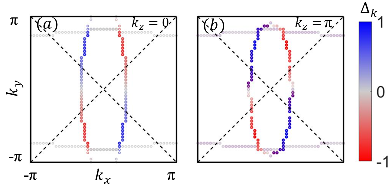}}
 	\end{center}
 	\caption{ (a), (b) The solution of the superconducting gap equation on the Fermi surface of the bulk BCO for two representatives ${k_z}$ values.  The pairing symmetry  in presence of inter-layer hybridization is $d_{xy}$ type with nodes along the $k_x=0$ and $k_y=0$ lines on the Fermi surface.   }
 	\label{ref:fig4}
 \end{figure}

The effect of inter-layer hybridization is even more significant for the ground-state superconducting gap functions. In Figs.~\ref{ref:fig4} (a) and (b) we plot the superconducting gap function of bulk BCO  for the largest pairing eigenvalues at two different $k_{z}$ values. The pairing symmetry of bulk BCO on the elliptical hole pocket can be expressed in the form $\Delta (\bf{k}) =  \Delta_{0} \sin \it (k_x) \sin (k_y)$ with line nodes along the $k_x=0$ and $k_y=0$ lines on the FS.  This is similar to the cuprate B$_{1g}$ superconducting basis function on a $\pi/4$ rotated axis. We find that although the sign of the gap remains unchanged along $k_z$, the gap function magnitude ($\Delta_{0}$) gets enhanced with increasing $k_{z}$. In bulk BCO, the pairing eigenvalue $\lambda$ seems to closely track the transition of unhybridized layer-I model (See SM Fig. S4). This feature is likely due to the dominance of the $d_{z^2}$ electrons and corresponding orbital resolved pairing interaction for the layer-I model. Our predictions can be probed by experimental techniques such as ARPES, STM among a variety of techniques that have been successfully utilized to understand the superconducting state in the cuprates.
	{\it Conclusions -}\label{discuss}
	The recent discovery of superconductivity in BCO at T$_c=73$K with high hole doping levels places this material in a new parameter regime among the various classes of cuprate high-temperature superconductors. 
 From the DFT calculations, we find that the $d_{z^{2}}$  orbital of the Cu(1) atom belonging to the octahedra lies at the Fermi level, and $d_{x^{2} -y^{2}}$ orbital is fully occupied. This makes BCO different from usual cuprate superconductors. We propose an effective 5-orbital tight-binding model consisting of the selective $d$- orbitals of the Cu atoms that shows excellent agreement with the DFT band structure when an inter-layer hybridization is included in the model. From the tight-binding analysis, we find the hybridization between the BCO layers, albeit weak, significantly influences the band structure along $Y\Gamma$ and $TZ$ where the van Hove singularities exist.  The $d$-orbitals coming from Cu(3)(/Cu(4)) and Cu(1) atoms form  electron like pocket at the Fermi level. Similar to the YBCO,  the planar layer of bulk BCO forms quasi-1D chain states. However, the hole pocket coming from $d_{x^{2} -y^{2}}$ orbital in YBCO\cite{16} and infinite layer nickelates\cite{17} are absent in BCO. 

In BCO, the presence of inter-layer hybridization plays a pivotal role in reshaping the Fermi surface. It removes the parallel Fermi pocket regions connected by dominant inter-orbital contributions. This leads to the dominance of intra-orbital nesting and susceptibility along the $(\pi-\delta,0)$ wave-vector [See SM  Section IV  for orbital resolved susceptibility contributions]. This diagonal intra-orbital pairing contribution will support an unconventional superconducting order. Additionally, the $(\pi-\delta,0)$ nesting wave-vector found in our susceptibility calculations would support $d_{xy}$ symmetry superconducting order.  We find an $s$-wave gap on Fermi pocket (without sign change over)  for layer-I, if we ignore the inter-layer hybridization. However, for layer-II, we do find a sign change of the superconducting gap over the Fermi pocket that is expected from repulsive interaction. Therefore, the gap over the entire Fermi pocket should not be considered as a conventional  $s$-wave gap but belonging to A$_{1g}$ symmetry with higher harmonic contributions that can lead to a sign change of gap on layer-II. The signatures of the nodal sign changing $d_{xy}$ gap can be probed in future thermodynamic measurements like low-temperature specific heat and  thermal conductivity measurements.   With synthesis of good quality single crystals, the gap should also be directly observable in ARPES and scanning tunneling spectroscopic experiments. 
 
 

 \textit{Acknowledgements:} This work was funded by the Department of Science and Technology, India, through Grant No. CRG/2020/004330. S.S. thanks SERB India for the Visiting Advanced Joint Research (VAJRA) program of the  Science and Engineering Research Board, Department of Science and Technology (SERB-DST), Government of India.
	
%

\hypersetup{colorlinks=true, urlcolor=blue, citecolor=blue}
\renewcommand{\thetable}{S\arabic{table}}
\renewcommand{\thefigure}{S\arabic{figure}}
\newcommand\x{\times}
\newcommand\bigzero{\makebox(0,0){\text{\huge0}}}
\newcommand*{\bord}{\multicolumn{1}{c|}{}}

\begin{widetext}
\pagebreak
\begin{center}
\textbf{Supplementary Materials for `` Unique $d_{xy}$ Superconducting State in the Cuprate Member Ba$_{2}$CuO$_{3.25}$"}
\end{center}

\section{Computational and Structural details} 
Density functional theory (DFT) calculations were performed using the Vienna ab-initio Simulation Package (VASP) \cite{vasp} code with projector-augmented-wave (PAW) \cite{PAW1, PAW2} pseudopotentials. Perdew–Burke–Ernzerhof (PBE) \cite{PBE} exchange-correlation functional scheme of generalized gradient approximation (GGA) was considered to take care of exchange and correlation functional. We have employed a kinetic energy cutoff of 500 eV and $\Gamma$-centred 16$\times$12$\times$4 $k$-mesh which yields 189 irreducible points for Brillouin zone (BZ) sampling. A Slater-Koster (SK) \cite{Slater} based minimal basis set tight-binding (TB) formalism is used to design the model Hamiltonian.
\renewcommand{\thefigure}{S1}
\begin{figure}[h]
\centering
\includegraphics[angle=-0.0,origin=0.5,scale=0.5]{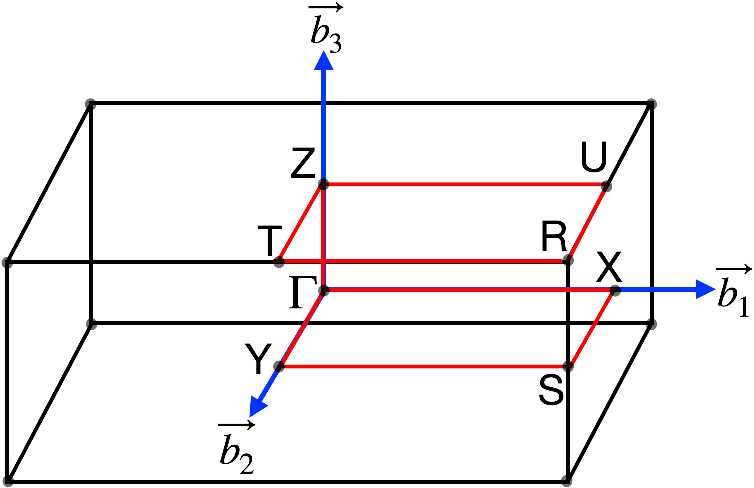}
\caption{ First Brillouin zone (BZ) of BCO unitcell crystal structure along with the high symmetry k-points.}
\label{ref:figS1}
\end{figure}
 
The crystal structure of Ba$_2$CuO$_{3.25}$ (BCO) is obtained from Ba$_2$CuO$_4$ by removing a few O-atoms as shown in Fig. \ref{ref:figS2} (a). The structure has been proposed in previous studies \cite{bco_dft1}. The primitive unitcell of BCO contains two inequivalent layers. As shown in Fig. \ref{ref:figS2} (a), layer-I forms an alternate CuO$_6$ octahedron, and CuO$_4$ square plane connected through the corner sharing O-atoms along the b-direction, and layer-II contains a one-dimensional chain of CuO$_4$ square plane extended along b-direction. Thus the unitcell has four inequivalent Cu atoms ( Cu(1), Cu(2) from layer-I and Cu(3), Cu(4) from layer-II ) and 13 O-atoms (seven in layer-I and six in layer-II).  
\renewcommand{\thefigure}{S2}
\begin{figure}
\centering
\includegraphics[angle=-0.0,origin=0.5,scale=0.6]{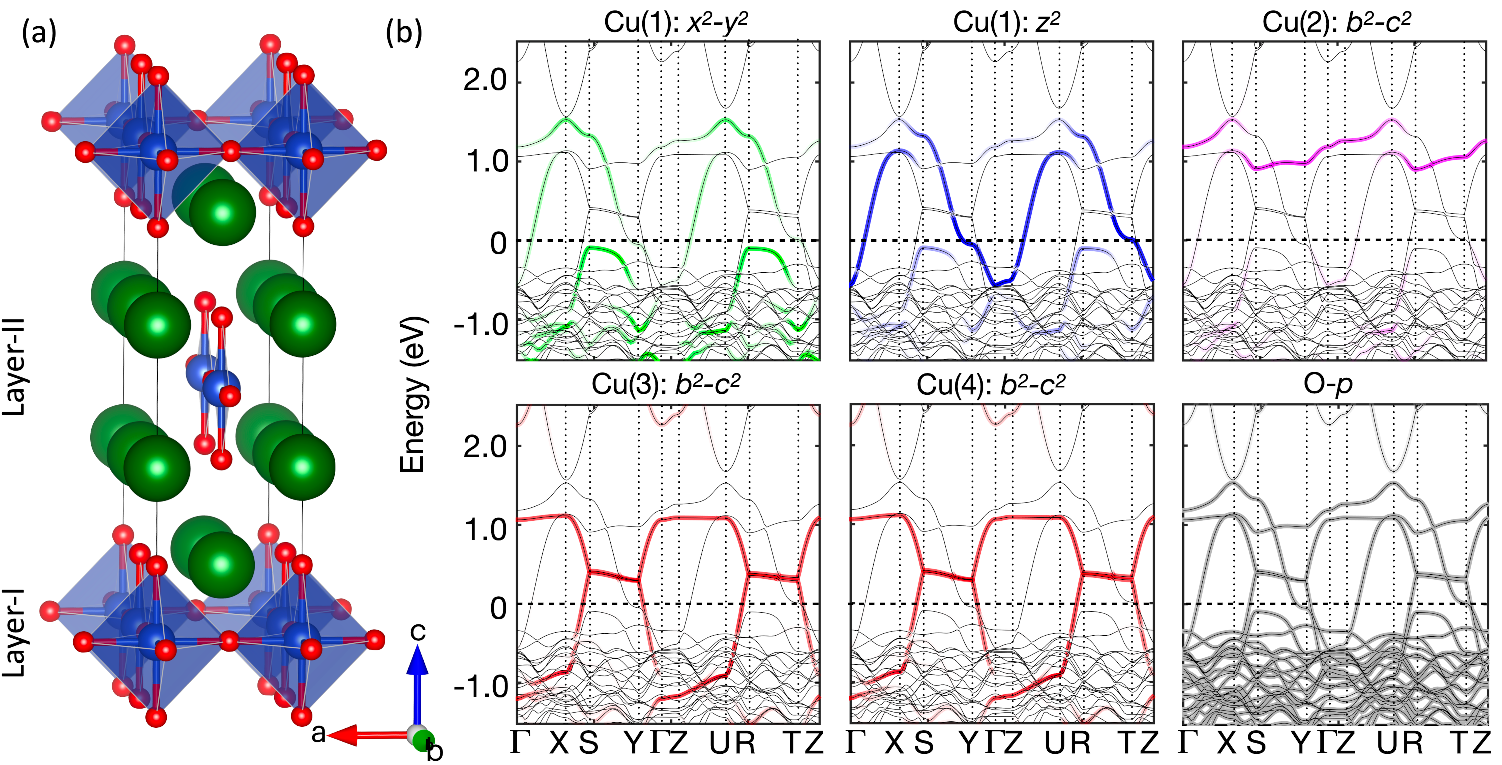}
\caption{ (a) Unitcell crystal structure of BCO. (b)  DFT obtained orbital resolved band structure of BCO. The bands are plotted along the high symmetry k-path provided in the BZ.
}
\label{ref:figS2}
\end{figure}
In order to get the correct orbital basis set required for constructing the model Hamiltonian, we have further performed the DFT calculations and studied the electronic structure properties of the compound. Fig. S2(b) shows the orbital resolved band structure of BCO, and it suggests that bands near the Fermi are contributed by the covalent hybridization of Cu-O atomic orbitals. However, not all hybridized orbitals contributions are present in the bands near the Fermi level, the exact molecular orbitals are different for different Cu atoms. For example, \{Cu(2), Cu(3), and Cu(4)\}-$x^2-y^2$ orbital hybridized with nearest-neighbor in-plane O-\{$p_x$, $p_y$\} orbitals, Cu(1)-$x^2-y^2$ hybridized with nearest-neighbor in-plane O-\{$p_x$, $p_y$\}, and Cu(1)-$z^2$ hybridized with nearest-neighbor out-of-plane O-$p_z$ orbital are contributing to the bands near the Fermi. Thus a total of 14 orbital basis-set with five Cu-$d$, and nine O-$p$ orbitals are sufficient to design the TB model Hamiltonian.
	
\section{Details of the tight-binding Hamiltonian }\label{rpasus}

 The matrix representation of the SK-TB model Hamiltonian of BCO is shown in Eq. 2 of the main text. Here, the sub-matrix $H$ of the layer-I in basis set order Cu(1) \{ $d_{z^2}$, $d_{x^2-y^2}$\}, Cu(2) \{$d_{x^2-y^2}$\}, $p_x$, $p_z$, $p_z$, $p_y$, and $p_y$ is:
 \renewcommand{\theequation}{1}
	\begin{equation}
		H_{l1} = \left(\begin{array}{cccccccc}
			\xi_{1} &  \xi_{1,2} &  \xi_{1,3} & \xi_{1,4} & \xi_{1,5} & 0 &  \xi_{1,7} & 0 \\ \cline{1-1}
			\bord & \xi_{2} &  \xi_{2,3} &  \xi_{2,4} & \xi_{2,5} & \xi_{2,5}^{*} & 0 & 0 \\ \cline{2-2}
			&\bord  &  \xi_{3} & 0 & \xi_{2,5}^{*} & \xi_{2,5} & 0 & 0 \\ \cline{3-3}
			& &\bord & \mu^{(3)} & 0 & 0 & 0 & 0 \\  \cline{4-4}
			& {\rm h.c.} & &\bord & \mu^{(1)} & 0 & 0 & 0 \\  \cline{5-5}
			&  & & & \bord & \mu^{(1)} & 0 & 0 \\  \cline{6-6}
			&  & & & &\bord & \mu^{(1)} & 0 \\  \cline{7-7}
			&  & & & & &\bord & \mu^{(1)} \\  \cline{8-8}
		\end{array} \right)
	\end{equation}
	h.c. denotes the Hermitian conjugate of the upper-triangular matrix.

		The Hamiltonian sub-matrix for layer-II in orbital basis set of Cu(3)/Cu(4) \{$d_{x^2-y^2}$\},  $p_z$, $p_z$, $p_y$ and $p_y$ is given as
 \renewcommand{\theequation}{2}	
 \begin{equation}
		H_{l2} = \left(
		\begin{array}{cccccc}
			\xi_{4} & \xi_{4,5} & \xi_{4,6} & 0 & 0 & 0 \\ \cline{1-1}
			\bord & \xi_{5} & 0 & \xi_{4,6} & 0 & 0 \\ \cline{2-2}
			&\bord & \mu^{(7)} & 0 & 0 & 0 \\ \cline{3-3}
			& {\rm h.c.} &\bord & \mu^{(7)} & 0 & 0 \\ \cline{4-4}
			& & &\bord & \mu^{(7)} & 0 \\ \cline{5-5}
			& & & &\bord & \mu^{(7)} \\ \cline{6-6}
		\end{array} \right)
	\end{equation}
	
		Further, the Hamiltonian sub-matrix containing the interaction between layer-I and layer-II is
   \renewcommand{\theequation}{3}
		\begin{equation}
		H_{l1-l2} = \left(
		\begin{array}{cccccc}
			0 & 0 & 0 & 0 & \xi_{12}^{2} & -(\xi_{12}^{2})^{*} \\
			0 & 0 & 0 & 0 & 0 & 0 \\
			\xi_{12}^{1} & (\xi_{12}^{1})^{*} & 0 & 0 & -(\xi_{12}^{2})^{*} & \xi_{12}^{2} \\
			0 & 0 & 0 & 0 & 0 & 0 \\
			0 & 0 & 0 & 0 & 0 & 0 \\
			0 & 0 & 0 & 0 & 0 & 0 \\
		\end{array} \right)
	\end{equation}
\renewcommand{\thefigure}{S3}
				\begin{figure}
		\begin{center}
			\rotatebox{0}{\includegraphics[width=0.6\textwidth]{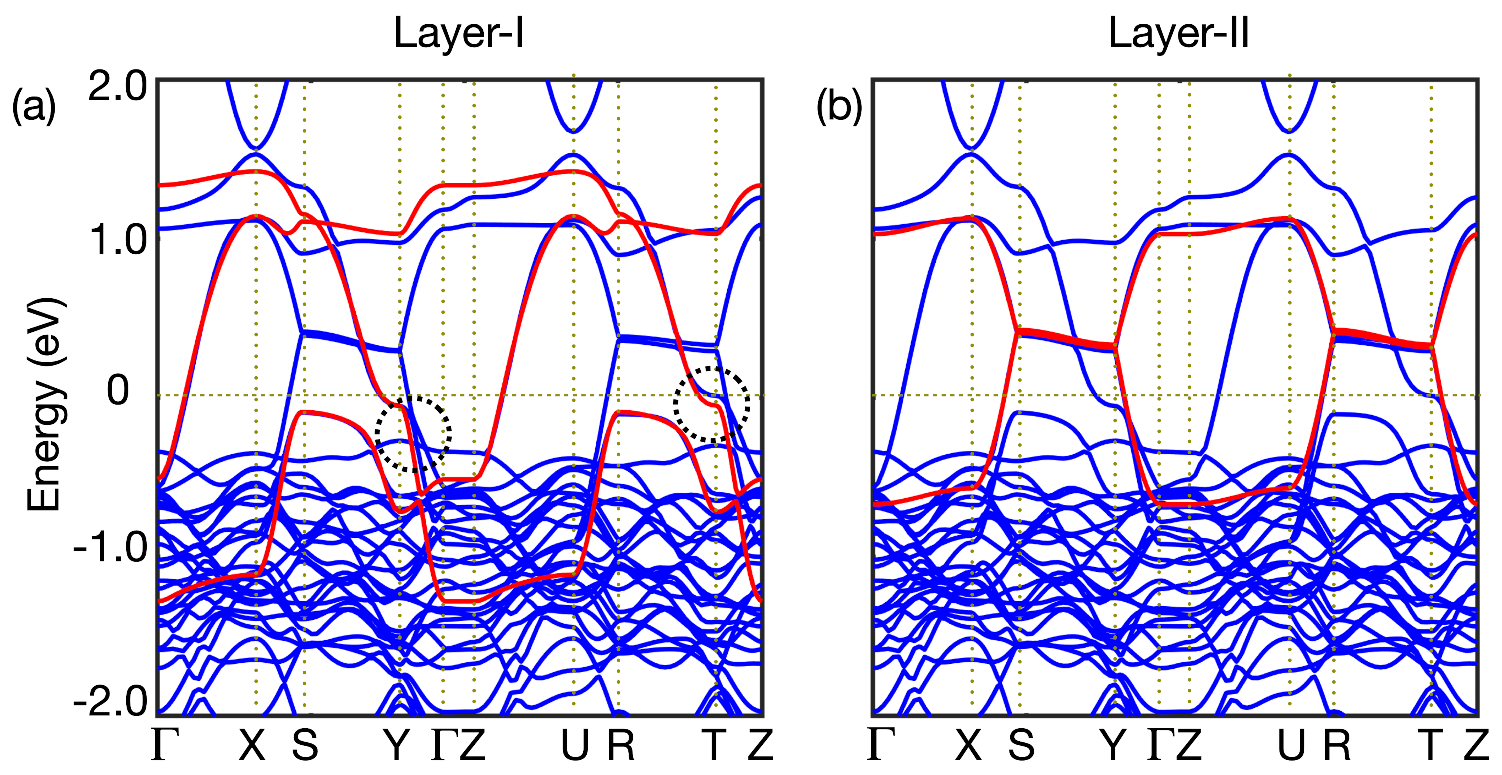}}
		\end{center}
		\caption{(Color online) (a) TB (red) fitted DFT (blue) band structure of layer-I of BCO crystal.  (b) Same for layer-II of BCO crystal. We plot bands along the high symmetry k-path in the BZ as shown in Fig. \ref{ref:figS1}. The dotted circles in (a) show the disagreement between DFT and TB bands at those regions when inter-layer interactions are ignored.
  }
		\label{ref:figS3}
	\end{figure}

 The components of the Hamiltonian matrices are found to be,
\renewcommand{\theequation}{4}
 \begin{subequations}
 	\begin{eqnarray}\nonumber
		\xi_{1}&=&t^{(5)}\cos(k_x) + \mu^{(4)} + t^{(15)}\cos(k_y)- 0.1\cos(2k_x)\\
		\xi_{2}&=&t^{(10)}\cos(ky) + t^{(11)}\cos(k_x)+ \mu^{(2)}\\
		\xi_{3}&=&\xi_{2}+\mu^{(5)}\\
		\xi_{4} &=&t^{(14)}\cos(k_x)+\mu^{(6)}\\
		\xi_{5}&=&\xi_{4}+0.02\\
		\xi_{1,2}&=&t^{(3)}\cos(k_x)+t^{(9)}\cos(k_y)\\
		\xi_{1,3}&=&t^{(4)}\exp(ik_y/4)\\
		\xi_{2,3}&=&t^{(1)}\cos(k_y/2)\\
		\xi_{1,4}&=&it^{(8)}\sin(k_x/2)\\
		\xi_{1,5}&=&it^{(8)}\sin(k_y/4)\\
		\xi_{1,7}&=&it^{(6)}\sin(0.156k_z)\\
		\xi_{2,4}&=&it^{(7)}\sin(k_x/2)\\
		\xi_{2,5}&=&t^{(4)}\exp(ik_y/4)\\
		\xi_{4,5}&=&t^{(12)}\cos(k_y/2)\\
		\xi_{4,6}&=&it^{(13)}\sin(k_y/4)\\
		\xi_{12}^{1}&=&2t_{12}^{(1)}\exp(-ik_y/4)\cos(k_z/2)\cos(k_x/2)\\
		\xi_{12}^{2}&=&it_{12}^{(2)}\exp(ik_y/4)\sin(k_z/3)\cos(kx/2)\\
	\end{eqnarray}
	 \end{subequations}
		The tight-binding parameters are, $t^{(1-14)}$ = $\big[$0.9775, 0.0237, 0.3396, 0.005, -0.78, 0.2076, -0.46 , 0.0, 0.035, -0.119, -0.074, 0.91, 0.6928, -0.054, -0.06.$\big]$
	$\mu^{(1-7)}$ = $\big[$ -0.74, -0.31, -1.24, 0.5, 1.18, 0.28, -1.64 $\big]$
	$t_{12}^{(1-2)}$ = $\big[$ -0.02, -0.3$\big]$. We further use Lo\"wdin downfolding mechanism to reduce the Hamiltonian size.
	The Lo\"wdin method used for the downfolding mechanism can be explained by,
 \renewcommand{\theequation}{5}
	\begin{equation}
		H^{\rm downfold}_{\alpha,\beta} = H_{\alpha,\beta} + \sum_{\gamma \neq \alpha}^{\prime} \frac{H_{\alpha,\gamma}(H_{\beta, \gamma})^*}{ H_{\alpha,\alpha} -H_{\gamma,\gamma}} 
	\end{equation}
	Here, $H^{\rm downfold}_{\alpha,\beta}$ is final 5$\times$5 downfolded Hamiltonian matrix, $\gamma$ contains O-$p$ -- O-$p$, Cu-$d$ -- O-$p$ orbitals interaction respectively which are projected on A.  We have numerically downfolded the full Hamiltonian and the resulting band structures are shown in Fig. \ref{ref:fig2} (a) of main text which shows an excellent agreement around the Fermi level with the all-electron band structure obtained from density functional theory (DFT) calculations.\\\\

	\section{Superconductivity}
	{\it RPA spin susceptibility-  }\label{rpasus} We use the multi-band Hubbard model to study the topology of Fermi surface (FS) and corresponding spin-fluctuation potential. The Hamiltonian of the Hubbard model is,

\renewcommand{\theequation}{6}
\begin{eqnarray} 
	H_{\rm int} &=& 
	\sum_{\alpha}\sum_{{\bf k,k',q}}U_{\alpha} c_{\alpha \uparrow}^{\dagger}({\bf k})c_{\alpha\downarrow}^{\dagger}({\bf  k}')c_{\alpha\downarrow}({\bf k'-q})c_{\alpha\uparrow}({\bf k+q}) \sum_{\substack{\alpha\ne \beta \\ \sigma\sigma'\in(\uparrow,\downarrow)}} \sum_{{\bf k,k',q}}\Big[ V_{\alpha,\beta} c_{{\bf k},\alpha,\sigma}^{\dagger}c_{{\bf  k}',\beta,\tilde{\sigma}}^{\dagger}c_{{\bf k'-q},\beta,\tilde{\sigma}}c_{{\bf k+q},\alpha,\sigma} \nonumber\\
	&&+\Big(V_{\alpha,\beta} -J_{H}\Big) c_{{\bf k},\alpha,\sigma}^{\dagger}  c_{{\bf  k}',\beta,{\sigma}}^{\dagger}c_{{\bf k'-q},\beta,{\sigma}}c_{{\bf k+q},\alpha,\sigma}\Big].
	\nonumber \\
	\label{Hint}
\end{eqnarray}
Here  $c^{\dagger}_{{\bf k},\alpha,\sigma}$ and $c_{{\bf k},\beta, \sigma}$ are fermion creation and annihilation operator in orbital $\alpha$ and $\beta$ and  $\tilde{\sigma}=-\sigma$.
Where $U_{\alpha}$ and $V_{\alpha,\beta}$ are the intra-orbital and inter-orbital Hubbard interaction between Cu-$d$ orbitals and $J_{H}$ is the Hund's coupling.

Non-interacting electron-hole density-density correlation function in the orbital basis is given by \cite{fwave,nickelates1} ,
\renewcommand{\theequation}{7}
\begin{eqnarray}
	[\chi_{0}(\textbf{q})]_{\alpha\beta}^{\gamma\delta}&=&-\frac{1}{N} \sum_{{\bf k},\nu\nu'}\psi^{\nu}_{\beta}({\bf k})\psi^{\nu\dagger}_{\alpha}({\bf k})\psi^{\nu'}_{\delta}({\bf k+q})\psi^{\nu'\dagger}_{\gamma} ({\bf k+q})\frac{f(\xi_{\nu'}({{\bf k+q}}))- f(\xi_{\nu}({{\bf k}}))}{\xi_{\nu'}({\bf k+q})-\xi_{\nu}({\bf k})+i\eta}.
	\label{Lindhard}
\end{eqnarray}

 N is the volume of the phase space. Using S-matrix expansion of the spin density and charge-density correlation function we obtain random-phase approximation (RPA) spin and charge susceptibilities,
 \renewcommand{\theequation}{8}
\begin{eqnarray}\label{spin_sus}
	\tilde{\chi}_{\rm s/c}({\bf q})= \tilde{\chi}_{0}({\bf q})\left(\tilde{\mathbb{I}} \mp \tilde{U}_{s/c}\tilde{\chi}_{0}({\bf q})\right)^{-1},
	\label{RPA}
\end{eqnarray}

 The nonzero components of onsite Hubbard interactions for spin and charge fluctuation are $\tilde{U}_{s}$ and $\tilde{U}_{c}$\cite{graser}. Nesting of the FS is captured in the Lindhard function. At the nesting vector, bare susceptibility shows a strong peak which leads to a large peak in the RPA spin susceptibility. The presence of $(1-\tilde{U}\chi_{0})$ in the denominator, contribution from the spin channel enhanced whereas the charge channel is suppressed due to $(1+\tilde{U}\chi_{0})$.  

	{\it Spin-fluctuation theory- }\label{rpasc}
We assume that the superconducting pairing in Cu-$d$ electrons is mediated via spin fluctuations.  We calculate the spin-fluctuation pairing potential by expanding the $ H_{\rm int}$ from Eq.~\eqref{Hint} into a perturbation series and collecting bubble and ladder diagrams. The effective Hamiltonian we obtain as \cite{fwave,nickelates1},
\renewcommand{\theequation}{9}
\begin{eqnarray}
	H_{\rm eff} &=& \sum_{\alpha\beta\gamma\delta}\sum_{{\bf kq},\sigma\sigma'} \Gamma_{\alpha\beta}^{\gamma\delta}({\bf q}) c_{\alpha \sigma}^{\dagger}({\bf k})c_{\beta\sigma'}^{\dagger}(-{\bf  k})c_{\gamma\sigma'}({\bf -k-q})c_{\delta\sigma}({\bf k+q}).
	\label{Hintpair}
\end{eqnarray}

 Here, the pairing potential is a tensor of four orbital indices. For singlet and triplet channels the spin-fluctuation pairing potential is given by \cite{SCrepulsive,SCcuprates,SCpnictides,SCHF,SCorganics},
 \renewcommand{\theequation}{10}
 \begin{subequations}
 	\begin{eqnarray}
 		\tilde{\Gamma}_{S}({\bf q})&=&\frac{1}{2}\big[3{\tilde U}_{s}{\tilde \chi}_{s}({\bf q}){\tilde U}_{s} - {\tilde U}_{c}{\tilde \chi}_{c}({\bf q}){\tilde U}_{c} + {\tilde U}_{s}+{\tilde U}_{c}\big],
 		\label{singlet}\\
 		\tilde{\Gamma}_{T}({\bf q})&=& -\frac{1}{2}\big[{\tilde U}_{s}{\tilde \chi}_{s}({\bf q}){\tilde U}_{s} + {\tilde U}_{c}{\tilde \chi}_{c}({\bf q}){\tilde U}_{c}\big].
 		\label{triplet}
 	\end{eqnarray}
 \end{subequations}

 Using unitary transformation we obtain pairing potential in the band basis. 
 \renewcommand{\thefigure}{S4}
 	\begin{figure}
		\begin{center}
		\includegraphics[width=0.4\textwidth]{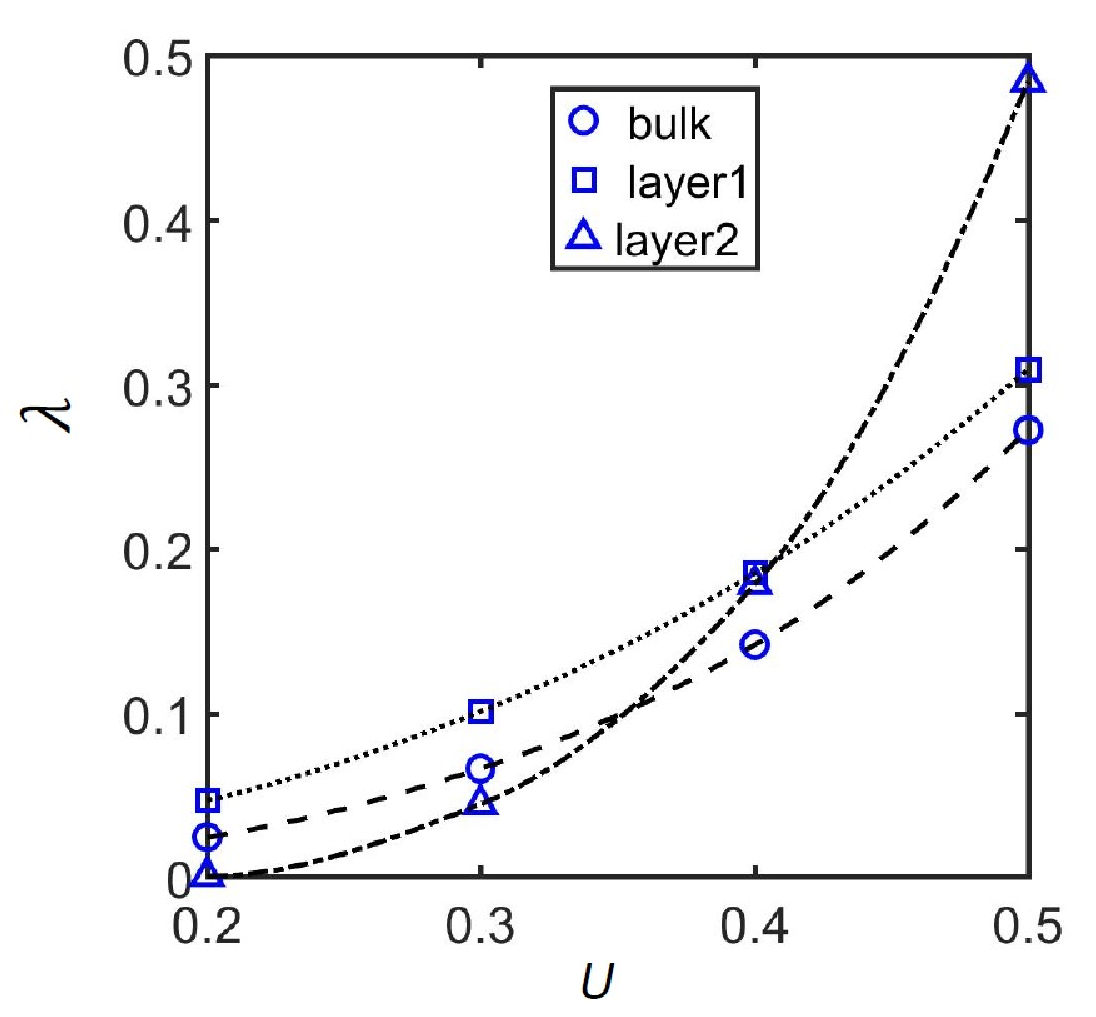}
		\end{center}
		\caption{We plot superconducting pairing eigenvalue as a function onsite Hubbard interaction for layer-I, layer-II, and bulk BCO.}
		\label{ref:figS4}
	\end{figure}



 \renewcommand{\thefigure}{S5}
\begin{figure}[b]
		\centering
  \rotatebox{0}{\includegraphics[width=1\textwidth]{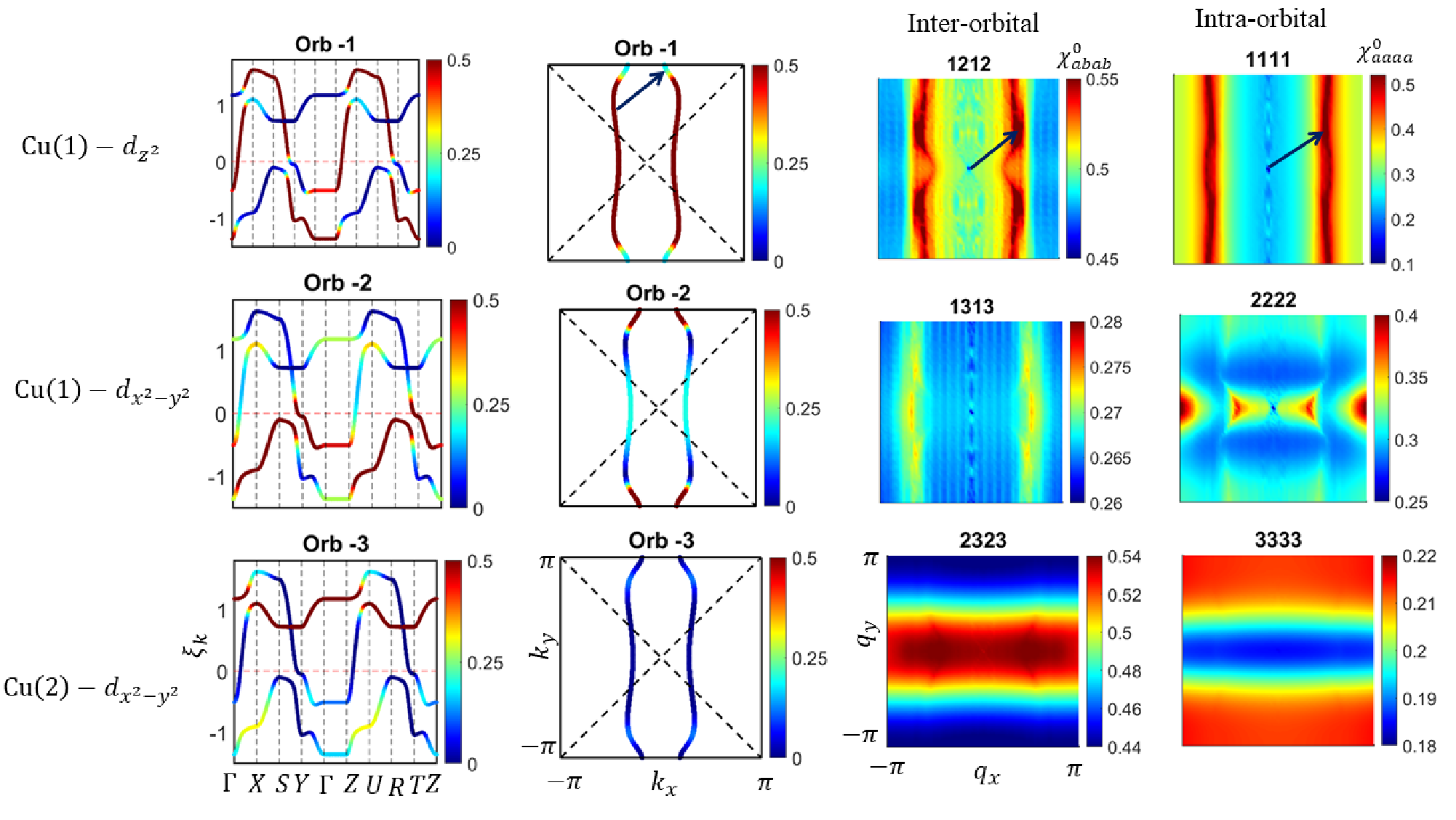}}
		\caption{ First column shows orbital weights on the bands of Layer-I in the absence of inter-layer hybridization, Second column shows orbital weights at the Fermi surface. Third and fourth column shows inter-orbital  and intra-orbital Lindhard  susceptibilities in absence of  the inter-layer hybridization. }
		\label{ref:figS5}
	\end{figure}

\renewcommand{\thefigure}{S6}
\begin{figure}
		\centering
  \rotatebox{0}{\includegraphics[width=1\textwidth]{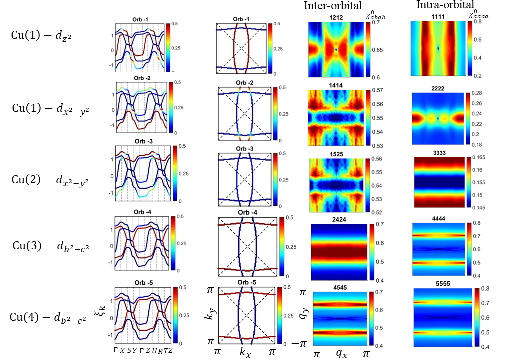}}
		\caption{ First column shows different orbital weights on BCO bands in the presence of inter-layer hybridization.  Second column shows orbital weights at the Fermi surface at $k_{z}=0$. Third and fourth col. shows inter and inter-orbital Lindhard susceptibilities respectively at $q_{z}=0$.  }
		\label{ref:figS6}
	\end{figure}

 \renewcommand{\theequation}{11}
 \begin{eqnarray}
 	\tilde{\Gamma}_{\mu\nu}({\bf k,q})&=&\sum_{\alpha\beta\gamma\delta} \Gamma_{\alpha\beta}^{\gamma\delta}({\bf q})  \psi^{\mu\dagger}_{\alpha}({\bf k})\psi^{\mu\dagger}_{\beta}(-{\bf k})\psi^{\nu}_{\gamma}({\bf -k-q})\psi^{\nu}_{\delta} ({\bf k+q}) \nonumber \\
 \end{eqnarray}
 We obtain superconducting pairing symmetry by solving the linearized gap equation,
 \renewcommand{\theequation}{12}
 	\begin{eqnarray}
 		\Delta_{\mu}({\bf k})= -\lambda\frac{1}{N}\sum_{\nu,{\bf q}}\tilde{\Gamma}_{\mu\nu}({\bf k,q})\Delta_{\nu}({\bf k+q}).
 		\label{SC2}
 	\end{eqnarray}
 $\lambda$ is known as a superconducting coupling constant. By solving  Eq. \eqref{SC2} we obtain pairing eigenfunction for largest eigenvalue. This largest eigenvalue determines the stability of superconducting gap function $\Delta({\rm \bf k})$\cite{SCrepulsive}.    
 
 The unconventional SC within spin fluctuation theory originates from the nesting at the FS. Since $\chi_{s}$ (see  Eq. \eqref{RPA}) is positive and larger than $\chi_{c}$, pairing potential Eq. \eqref{singlet} is repulsive. The only possible solution of the gap equation for repulsive interaction is when $\Delta$ (see  Eq. \eqref{SC2} ) changes sign between momentum vectors ${\bf k}$ and ${\bf k + Q}$, where ${\bf Q}$ is the nesting vector that connects the Cooper pairs. This leads to an anisotropic solution of the gap function in the momentum space which should reflect underlying point group symmetry. \par
 We study the effect of Hubbard interaction on the superconducting pairing eigenvalue (see Fig.~\ref{ref:figS4} ). This result shows a comparison of dominant pairing eigenvalue for layer-I, and layer-II for unhybridized bands and bulk BCO with hybridized bands as a function of onsite Hubbard interaction. We see that pairing eigenvalue $\lambda$ extracted by solving the linearized gap equation is slightly larger on layer-II in comparison to layer-I for the unhybridized bands for $U>0.4$. However, the decoupled quasi-1D layer would generically be more susceptible to quantum fluctuations that are likely to suppress the transition temperature strongly.

\section{Orbital resolved susceptibilities- The role of interlayer hybridization }

\subsubsection*{Absence of inter-layer hybridization}

In Fig.~\ref{ref:figS5}, we have  illustrated the orbital resolved susceptibilities and electronic structures when inter-layer hybridization is absent. The key insight provided by Fig.~\ref{ref:figS5} lies in the comparison between the inter and intra-orbital susceptibilities shown in the  third and fourth columns. Specifically, we have demonstrated that the inter-orbital susceptibility is dominant over the intra-orbital susceptibility. This observation is essential in understanding how attractive pairing interactions can arise without Hund's coupling.

The physical susceptibility in the main manuscript suggests that intra-orbital susceptibility might dominate due to $(\pi-\delta,0)$ nesting, but a closer look (Fig.~\ref{ref:figS5}) reveals that inter-orbital susceptibility is actually dominant. This inter-orbital susceptibility peaks at a different nesting vector and is smaller at $(\pi-\delta,0)$. This finding explains why layer-I can create an effective attractive pairing channel.

\subsubsection*{Presence of inter-layer hybridization}
In Fig.~\ref{ref:figS6}, we have shown the orbital weights across different bands and the Fermi surface of BCO under the influence of inter-layer hybridization.  As expected, the presence of inter-layer hybridization significantly alters the Fermi surface and orbital weights near the $(0, \pi)$ region, thereby makes a significant effect on  the pairing interactions.

Specifically, as can be seen from the calculated orbital susceptibility plots, the introduction of inter-layer hybridization results in a reduction of the contribution from inter-orbital susceptibility. This change in susceptibility is a direct consequence of the modified Fermi surface and orbital weights. In particular, in the presence of hybridization, the elliptical Fermi pocket for layer-I is now completely dominated by the $d_{z^2}$ orbital that will now support susceptibility in intra-orbital channel. Secondly, the modification of the Fermi pocket shape near the hybridization region suppresses the parallel Fermi pocket regions that previously supported the inter-orbital nesting in the absence of inter-layer hybridization. As is well known, such parallel regions are required to have sufficient weight from the $\bf k$-sum at a given {\bf  q} wave-vector in susceptibility calculations. 
The above scenario therefore not only will support a nesting vector along the $(\pi-\delta, 0)$ wave-vector, but the susceptibility will be driven by the dominant intra-orbital contributions. This will naturally lead to a dominant intra-orbital pairing (or dominant diagonal contribution to the pairing matrix) and the corresponding repulsive pairing interaction plays a pivotal role in the emergence of the observed $d_{xy}$ pairing symmetry in BCO.

	\vspace{1cm}
\end{widetext}

\end{document}